\documentclass[11pt]{article}
\usepackage{amsmath,amsfonts}

\begin{document}
\thispagestyle{empty}

\begin{center}
\Large
A Bubble Theorem
\end{center}

\vspace*{0.2 in}
 
\begin{flushright}
Oscar Bolina  \\
University of California \\
Davis, CA 95616-8633\\
bolina@math.ucdavis.edu
\vskip .25 in               
J. Rodrigo Parreira \\        
Cluster Consulting \\
Torre Mapfre pl 38 \\
Barcelona, 080050 Spain

\end{flushright}

\vskip .25 in

\paragraph*{Introduction}

It is always a good practice to provide the physical content of an
analytical result. The following algebraic inequality lends itself
well to this purpose: For any finite sequence of real numbers $r_{1},
r_{2}, ..., r_{N}$, we have
$$
(r_{1}^{3}+r_{2}^{3}+ ... + r_{N}^{3})^{2} \leq
(r_{1}^{2}+r_{2}^{2}+ ... + r_{N}^{2})^{3}. \eqno(1)
$$
A standard proof is given in \cite{HLP}. An alternative proof follows
from the isoperimetric inequality
$$
\label{IP2}
A^{3} \geq 36 \pi V^{2},
$$
where {\it A} is the surface area and {\it V} the volume of any
three-dimensional body. Setting the area $A=\sum_{i=1}^{N} 4 \pi r^{2}$
and the volume $V=\sum_{i=1}^{N} (4/3) \pi r^{3}$ yields (1).

\paragraph*{A Bubble Proof}
We give yet another proof, now using elements of surface tension
theory and ideal gas laws to the formation and coalescence of 
bubbles. This proof, found in \cite{B}, runs as follows.
\newline
According to a well-known result in surface tension theory, when a
spherical bubble of radius {\it R} is formed in the air, there is a
difference of pressure between the inside and the outside of the surface
film given by 
$$
p=p_{0} + \frac{2T}{R}, \eqno(2)
$$
where $p_{0}$ is the (external) atmospheric pressure on the surface
film of the bubble, {\it p} is the internal pressure, and {\it T} is
the surface tension that maintains the bubble \cite{A}.

Suppose initially that {\it N} spherical bubbles of radii $R_{1},
R_{2}, ... , R_{N}$ float in the air under the same surface
tension {\it T} and internal pressures $p_{1}, p_{2}, ... p_{N}$.
According to (2),
$$
p_{k}=p_{0} + \frac{2T}{R_{k}}, \;\;\;\;\;\;\;\;\;\;\;\;\;\;\; k=1,2, ...
N. \eqno(3)
$$
Now suppose that all {\it N} bubbles come close enough to be
drawn together by surface tension and combine to form a single  
spherical bubble of radius {\it R} and internal pressure {\it p},  
also obeying Eq. (2). When this happens, the product of the internal
pressure {\it p} and the volume {\it v} of the resulting bubble 
formed by the coalescence of the initial bubbles is, according to 
the ideal gas law \cite{A}, given by
$$
pv=p_{1}v_{1}+...+p_{N}v_{N}, \eqno(4)
$$
where $v_{k}$ (k=1,2,..., N) are the volumes of the individual 
bubbles before the coalescence took place. For spherical bubbles, 
(4) becomes
$$
pR^{3}=p_{1}R_{1}^{3}+...+p_{N}R_{N}^{3}. \eqno(5)
$$
Substituting the values of {\it p} and $p_{k}$ given in
(2) and (3) into (5), we obtain
$$
R^{3}-R_{1}^{3}-R_{2}^{3}- ... - R_{N}^{3} =\frac{2 T}{p_{0}}
(R_{1}^{2}+R_{2}^{2}+ ... + R_{N}^{2}-R^{2}). \eqno(6)
$$
Now, if the total amount of air in the bubbles does not change, the
surface area of the resulting bubble formed by the coalescence of the
bubbles is always smaller than the sum of the surface area of the
individual bubbles before coalescence. Thus,
$$
R_{1}^{2}+R_{2}^{2}+ ... + R_{N}^{2} \geq R^{2}.\eqno(7)
$$
Since the potential energy of a bubble is proportional to its
surface area, (7) is a physical condition that the surface
energy of the system is minimal after the coalescence.

It follows from (7) and the fact that $p_{0}$ and {\it T} are
positive constants that the left hand side of equation (6)
satisfies
$$
R_{1}^{3}+R_{2}^{3}+ ... + R_{N}^{3} \leq R^{3}. \eqno(8)
$$
The equality, which implies conservation of volumes, holds
when the excess pressure in the bubble film is much less the
atmospheric pressure. Combining (7) and (8) yields the 
inequality (1), which is also valid for negative numbers.\\

{\small 
\paragraph*{Acknowledgment.}    
O.B. would like to thank Dr. Joel Hass for pointing out the
isoperimetric proof of (1), and FAPESP for support under 
grant 97/14430-2.
}

\end{document}